\documentclass[aps,prl,twocolumn,showpacs]{revtex4}
\usepackage[T1]{fontenc}
\usepackage[latin1]{inputenc}
\usepackage{graphicx}
\usepackage{hyperref}
\usepackage{amsmath}

\begin{document}

\title{$q$-Breathers in Discrete Nonlinear Schr\"{o}dinger arrays with weak disorder}
\author{M.~V.~Ivanchenko}

\affiliation{Department of Applied Mathematics, University of Leeds, LS2 9JT, Leeds, United Kingdom}

\begin{abstract}
Nonlinearity and disorder are key players in vibrational lattice dynamics, responsible for localization and delocalization phenomena. $q$-Breathers -- periodic orbits in nonlinear lattices, exponentially localized in the reciprocal linear mode space -- is a fundamental class of nonlinear oscillatory modes, currently found in disorder-free systems. In this paper we generalize the concept of $q$-breathers to the case of weak disorder, taking the Discrete Nonlinear Schr\"{o}dinger chain as an example. We show that $q$-breathers retain exponential localization near the central mode, provided that disorder is sufficiently small. We analyze statistical properties of the instability threshold and uncover its sensitive dependence on a particular realization. Remarkably, the threshold can be intentionally increased or decreased by specifically arranged inhomogeneities. This effect allows us to formulate an approach to controlling the energy flow between the modes. The relevance to other model arrays and experiments with miniature mechanical lattices, light and matter waves propagation in optical potentials is discussed.
\end{abstract}

\pacs {63.20.Pw, 63.20.Ry, 05.45.-a }

\maketitle

A wealth of physical systems from natural crystals to the cutting-edge technology products like micro and
nanomechanical system arrays are spatially extended and discrete. Interaction between their
elements is a key source for a number of fundamental dynamical and statistical physical phenomena including thermal
conductivity, wave excitation and propagation, electron and phonon scattering. To provide with a full understanding of these processes the theory of collective vibrational modes is in demand. The principal question to be answered is the effect of the two fundamental
features of any lattice: nonlinearity and disorder. 

In recent decades we have witnessed a remarkable progress in studying their impacts separately. Nonlinearity induces interaction between linear normal modes and energy sharing if it is strong enough (the Fermi-Pasta-Ulam (FPU) problem) \cite{fpu,Ford}, or exponential localization of exact periodic solutions (discrete breathers) in space \cite{flach_review}. Independently, disorder leads to exponentially localized linear vibrational modes (Anderson modes) \cite{anderson}. 

However, the concurrent effect of nonlinearity and disorder has not received a satisfactory full description yet. \emph{Strongly disordered and weakly nonlinear} lattices enjoy intensive research, in particular, on continuation of Anderson modes into nonlinear regime \cite{anderson_continuation}, wavepacket spreading \cite{spreading}, light propagation in photonic lattices \cite{segev}, and Bose-Einstein condensate (BEC) localization in random optical potentials \cite{lahini}. In contrast, little is known in case of \emph{pronounced nonlinearity and weak disorder}. Importantly, this situation is realized in micro and nano-mechanical oscillatory arrays that are often driven into nonlinear regime, while the spatial disorder is constantly reduced by improving fabrication techniques \cite{Sievers,Roukes2}. On the atomic scale, the surface vibrational modes are thought to be a source of selective catalytic properties of three-dimensional gold nano-clusters for a variety of chemical reactions \cite{gold_clusters}.  Light propagation and BEC dynamics in random optical media are equally strong motivating problems.

$q$-Breathers (QBs) present a recently discovered fundamental class of nonlinear oscillatory modes. They are exact time-periodic solutions to nonlinear lattice equations, continued from linear normal modes and exponentially localized in the linear mode space. Introduced to explain the FPU paradox (energy locking in the low-frequency part of the spectrum, recurrencies, and size-dependent stochasticity thresholds) \cite{we_qb}, they have been found in two and three dimensional FPU arrays and discrete nonlinear Schr\"{o}dinger (DNLS) lattices \cite{we_dnls}; last year quantum QBs were observed in the Bose-Hubbard chain \cite{qqb}. QBs have been suggested as key actors in a BEC pulsating instability \cite{qb_in_bec} and a four-wave mixing process in a nonlinear crystal \cite{qb_in_crystal}. 

In this paper we extend the concept of $q$-breathers to random arrays, implementing the paradigmatic DNLS model as an example. The cornerstones of our approach are continuation of QBs into non-zero 'frozen' disorder, taking a nonlinear localized solution as a seed, and analysing statistics then. We show that QBs display the crossover from the exponential localization near the central mode to the power-law decay at a distance. Their average linear stability  threshold in nonlinearity keeps the same value in the first order approximation. The variance increases linearly with disorder, manifesting high sensitivity on particular realizations. Finally, we demonstrate, that the superimposed periodic modulation of the linear coupling strength can be a means of the energy flow control.

The DNLS lattice is represented by the Hamiltonian
\begin{equation}\label{eq6}
\begin{aligned}
  &H=\sum\limits_{n}
  ((1+D\kappa_n)\psi_{n-1}\psi_{n}^{*}+(1+D\kappa_{n+1})\psi_{n+1}\psi_{n}^{*}+\\
  &+\frac{\mu}{2}|\psi_n|^4),
  \end{aligned}
\end{equation}
and the equations of motion are $i\dot{\psi_n}=\partial H/\partial\psi_{n}^{*}$:
\begin{equation}\label{eq6a}
i\dot{\psi_n}=(1+D\kappa_n)\psi_{n-1}+(1+D\kappa_{n+1})\psi_{n+1}+\mu\left|\psi_n\right|^2\psi_n
\end{equation}
 Here $\psi$ is a complex scalar that may describe the slow small-amplitude dynamics of a classical nonlinear oscillators array \cite{kivshar,johansson}, probability amplitude of an atomic cloud on an optical lattice
site \cite{healinglength}, or the amplitudes of a propagating
electromagnetic wave in an optical waveguide \cite{nonlinearoptics}. Zero boundary conditions apply: $\psi_0=\psi_{N+1}=0$. $\mu$ and $D$ are the nonlinearity and disorder parameters, random $\kappa_{n}\in[-1/2,1/2]$ are uniformly distributed and uncorrelated: $\left\langle \kappa_{n}\kappa_{m}\right\rangle=\sigma^2_{\kappa}\delta_{n,m}$, $\sigma^2_\kappa=1/12$.
Beside the total energy, the norm $B=\sum\limits_{n} |\psi_{n}|^2$ is conserved. Changing $\mu$ is strictly equivalent to changing
the norm $B$, thus we fix $B=1$ further on.
The canonical transformation to the reciprocal space of normal
modes with new variables $Q_{q}(t)$ is given by
\begin{equation}\label{eq7}
\psi_{n}(t)=\sqrt{\frac{2}{N+1}}\sum\limits_{q}^N Q_{q}(t)\sin{\frac{\pi q n}{N+1}},
\end{equation}
and the dynamics in this space reads:
\begin{equation}\label{eq8}
\begin{aligned}
 & i\dot{Q}_q+\Omega_{q}Q_{q}=
\frac{\rho}{2}\sum\limits_{p,r,s} G_{q,p,r,s}
  Q_{p}Q_{r}Q_{s}^*+d\sum\limits_p V_{q,p}Q_p,
  \end{aligned}
\end{equation}
where $\rho=\frac{\mu}{N+1}, \ d=\frac{D}{\sqrt{N+1}}$, $\Omega_{q}=-2\cos{\frac{\pi q}{N+1}}$ are the normal mode frequencies for the
linear disorder-free system with $\mu=D=0$. The nonlinear intermode coupling coefficients are $G_{q,p,r,s}=\sum\limits_{\pm}
  (-1)^{(\pm p)(\pm r)(\pm s)} \left(\delta_{q\pm p \pm r \pm s,0}
  +\delta_{q\pm p \pm r \pm s,\pm 2(N+1)}\right)$ and the disorder induced ones read
$V_{q,p}=\frac{2}{\sqrt{N+1}}\times\sum\limits_{n=1}^{N-1}\kappa_n(\sin\frac{\pi q n}{N+1}\sin\frac{\pi p (n+1)}{N+1}+\sin\frac{\pi q (n+1)}{N+1}\sin\frac{\pi p n}{N+1})$.

In the disorder-free case QBs are time-periodic stationary solutions $\psi_{n}(t)=\phi_{n} \exp(i\Omega t)$ with the frequency $\Omega$ and time-independent amplitudes $\phi_{n}$ localized in normal mode space. In the $q$-space they have the form
$Q_{q}(t)=A_{q}\exp(i\Omega t)$, the amplitudes of the modes $A_{q}$ being time-independent
and related to the real-space amplitudes by the canonical transformation (\ref{eq7}); the mode energies are defined as $B_q=\left|A_q\right|^2$.
Here we focus on time-reversible periodic orbits and, thus, consider $A_q$ to be real numbers. The amplitudes satisfy a closed system of algebraic equations:
\begin{equation}\label{eq9}
\left\{
\begin{aligned}
&(\Omega_{q}-\Omega) A_{q}=\frac{\rho}{2}
\sum\limits_{p,r,s}
G_{q,p,r,s} A_{p}A_{r}A_{s}^*+d\sum\limits_p V_{q,p}A_p,\\
&\sum\limits_{q} \left|A_{q}\right|^2 - B = 0
\end{aligned}
\right.
\end{equation}

Our methodology consists of two steps. Firstly, we take a known QB solution for non-zero nonlinearity \cite{we_dnls}. A particular realization of $\{\kappa_n\}$ is chosen and $d$ regarded as the disorder parameter. Together with the nonlinearity parameter $\rho$, it is assumed to be small $\rho, d\ll 1$. Then, an asymptotic expansion in powers of $\{\rho, d\}$ is developed. Subsequent linear stability analysis employs the constructed solution. Secondly, statistical properties of the QB solution and the instability threshold  are analyzed.

Continuation of QBs from $\mu\neq0, D=0$ to $\mu,D\neq 0$ exploits the same ideas as from $\mu=D=0$ to $\mu\neq 0,D=0$ \cite{we_dnls}. 
For small amplitude excitations the nonlinear and disorder terms in
(\ref{eq8}) can be neglected and the $q$-oscillators get decoupled, their
harmonic energy $B_{q}=\left|Q_{q}\right|^2$ being conserved in time. Single $q$-oscillator excitations
($B_{q} \neq 0$ for $q\equiv q_0$ only) are trivial stationary and $q$-localized solutions for $\beta=D=0$.

In the disorder-free case such periodic orbits can
be continued into the nonlinear case at fixed total energy \cite{we_dnls} by solving the system of algebraic equations (\ref{eq9}), granted by the implicit function theorem \cite{implicit},
as the non-resonance condition $\Omega_{q_0} \neq \Omega_{q \neq q_0}$ holds. This is valid for $d\ll 1$ as well, for the spectrum remains non-resonant with the probability $1$ \cite{anderson_continuation}. Numerically, we were able to continue QBs into the ${\beta,D}\neq 0$ domain for all parameters taken. 

\begin{figure}[t]
{\centering
\resizebox*{0.95\columnwidth}{!}{\includegraphics{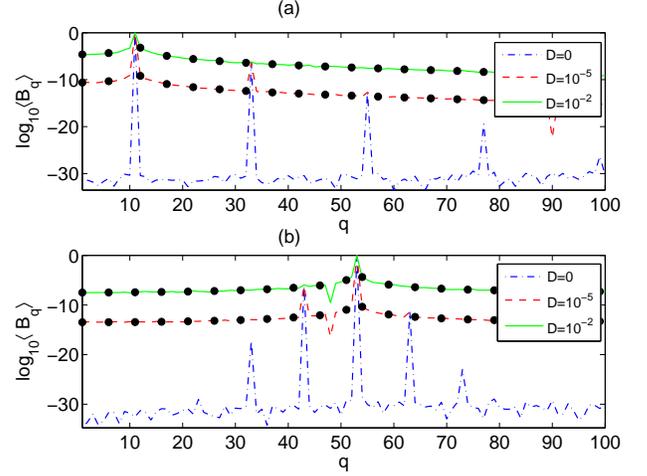}}}
{\caption{The average mode energy distribution in QBs with increase of disorder, where
$\mu=0.1, N=100$ : (a) the low frequency mode $q_0=11$ and (b) the middle frequency mode $q_0=53$. Filled circles are analytical estimates (\ref{eq11})}\label{fig5}}
\end{figure}

Typical results for the low-frequency and middle-frequency QBs are shown in Fig.\ref{fig5}. They demonstrate the crossover between the exponential localization and the disorder induced background. The disorder-free exponential localization persists in some neighborhood of the central mode for sufficiently small disorder, but is range shrinks as disorder grows. High-frequency QBs behave analogously.

Let us construct an asymptotic expansion for the QB solution. We assume $\rho, d \ll 1$ and start from the disorder-free QB profile $A^{NL}_q$ for the modes $q_0$, $3q_0$,\dots,$(2n+1)q_0$,\dots$\ll N$ in the leading order of $\rho$ \cite{we_dnls}: 
\begin{equation}\label{eq10b}
\begin{aligned}
&A^{NL}_{(2n+1)q_0}=(-1)^n\gamma^n A_{q_0}, \ \gamma=\frac{\mu(N+1)}{16\pi^2q^2}B_{q_0}, \\
&\Omega^{NL}=\Omega_{q_0}-\frac{\rho}{2}A_{q_0}^2 
\end{aligned}
\end{equation}
We seek an asymptotic expansion in powers of $d\ll 1$: $\hat{A}_q=A^{(0)}_q+d A^{(1)}_q+O(d^2,\rho d), \ \hat{\Omega}=\Omega^{(0)}+d \Omega^{(1)}+O(d^2,\rho d)$, where $A^{(0)}_q=A^{NL}_q, \ \Omega^{(0)}_q=\Omega^{NL}_q$. Substitution into (\ref{eq9}) gives 
\begin{equation}\label{eq10c}
A_q^{(1)}=\frac{V_{q,q_0}}{\Omega_q-\Omega_{q_0}}A_{q_0}, \ q\neq q_0, \ \Omega^{(1)}=-V_{q_0,q_0}
\end{equation}
 The ensemble average of the "disorder contribution" to the energy $B_q^{DO}=\left|d A_q^{(1)}\right|^2$ is 
\begin{equation} 
\label{eq11}
\left\langle B_q^{DO}\right\rangle=\frac{2 d^2 \sigma^2_\kappa(1+\Omega_q\Omega_{q_0}/4)}{(\Omega_q-\Omega_{q_0})^2}B_{q_0}\;,
\end{equation}
that approximates well the disorder-dominated part of the numerically obtained QB profiles in different parts of the linear spectrum (Fig.\ref{fig5}). The power-law decay $\left\langle B_q^{DO}\right\rangle\propto (q-q_0)^{-2}$ fits in the large part of the $q$-space. One can estimate the crossover location between the exponential decay and the power-law, in particular, when the modes next the to the central one become excited almost equally well. Letting $q=q_0+1$ we obtain the "small" $\sigma_\kappa D\ll \pi/(2(N+1)^{3/2})$ and the "large" $\sigma_\kappa D\gg \pi/\sqrt{(N+1)}$ disorder criteria for delocalization of the least robust modes $q_0=1,N$ and the most robust one $q_0=N/2$. Thus, single-cite centered modes do not exist above the size-dependent threshold in disorder magnitude. (Note, that delocalization in the mode space approximately corresponds the onset of the Anderson localization in the direct space.) 

The linear stability of QBs is determined by considering the evolution of small complex-valued perturbations $\zeta_q(t)$ to the stationary solution \cite{we_dnls}: $Q_q(t)=(\hat{A}_q+\zeta_q(t)) \exp{(i \Omega t)}$. In linearized equations the stability requires all the eigenvalues be negative. Numerically we solve the correspondent problem in the direct space (\ref{eq6a}). In the following we restrict our attention to the low and middle-frequency QBs, leaving the more complex case of $q_0>N/2$ (when for $D=0$ the instability threshold behaves erratically vs. $q_0$ \cite{we_dnls}) for the future study. 

\begin{figure}[t]
{\centering
\resizebox*{0.95\columnwidth}{!}{\includegraphics{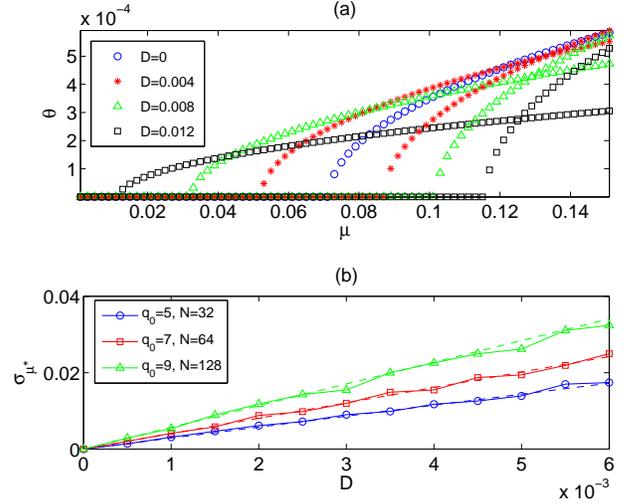}}}
{\caption{(a) The maximal eigenvalues $\theta$ of QBs with $q_0=15, N=128$ and two different sets of $\{\kappa_n\}$ vs. the nonlinearity coefficient for several values of disorder strength $D$. For one realization of disorder the instability threshold is increasing with $D$, for another -- decreasing. (b) The variance of the QB instability threshold $\sigma_{\mu^*}$ vs. $D$. Solid lines are analytical estimates (\ref{eq11})}\label{fig6}}
\end{figure}

We find, that the instability develops similarly for zero and non-zero disorder, the increase or decrease of the threshold $\mu^*$ sensitively depending on a particular realization (Fig.\ref{fig6}). The average $\left\langle \mu^*\right\rangle$ remains very close to the zero-disorder value $\mu_0^*$. In contrast, the variance $\sigma_{\mu^*}$ is significantly growing, depending on $D$ almost linearly (Fig.\ref{fig6}, deviations being observed when the probability of $\mu^*$ being next to zero becomes substantial).   

The analytic study of the QB stability has not been done before (even for $D=0$) and we present it here for the first time (restricting to $q_0<N/2$ as above). Linearized equations for small perturbations read:
\begin{equation}\label{eq10}
\begin{aligned}
&i\dot\zeta_q=(\hat{\Omega}-\Omega_q)\zeta_q+\frac{\rho}{2} B_{q_0}\sum G_{q,q_0,q_0,p}(\zeta_p^*+2\zeta_p)\\
&+d\sum V_{q,p}\zeta_p
\end{aligned}
\end{equation}

In analogy to the FPU chain \cite{we_qb} we suggest (and verify that by comparison with the numerical results) that the eigenvectors for the main instability be almost parallel to the subspace $\{\zeta_q=0: \ q\neq q_0\pm1\}$. Thus we arrive at a simpler task of finding eigenvalues of the system of two complex-valued linear equations (retaining $O(\rho,d)$ terms only):
\begin{equation}\label{eq10a}
\begin{aligned}
&i\dot\zeta_{q_0-1}=(\hat{\Omega}-\Omega_{q_0-1}+d V_{q_0-1,q_0-1})\zeta_{q_0-1}+\\
&+\frac{1}{2}\rho B_{q_0}(\zeta_{q_0+1}^*+2\zeta_{q_0+1})+d V_{q_0-1,q_0+1}\zeta_{q_0+1},\\
&i\dot\zeta_{q_0+1}=(\hat{\Omega}-\Omega_{q_0+1}+d V_{q_0+1,q_0+1})\zeta_{q_0+1}+\\
&+\frac{1}{2}\rho B_{q_0}(\zeta_{q_0-1}^*+2\zeta_{q_0-1})+d V_{q_0-1,q_0+1}\zeta_{q_0-1}
\end{aligned}
\end{equation}
 After an extensive algebra one finally gets the bifurcation point:
\begin{equation}
\label{eq11}
\begin{aligned}
&\mu^*\approx\mu^*_0\left(1-\frac{d}{\pi^2\left|\Omega_{q_0}\right|}\Delta V_{q_0,q_0}\right),\\
&\left\langle \mu^*\right\rangle\approx\mu^*_0, \ \sigma_{\mu^*}\approx\frac{D\sigma_\kappa\sqrt{3(N+1)}}{B_{q_0}}, 
\end{aligned}
\end{equation}
where $\Delta V_{q_0,q_0}=V_{q_0-1,q_0-1}-2V_{q_0,q_0}+V_{q_0+1,q_0+1}$, and the disorder-free $\mu^*_0=\frac{\pi^2\left|\Omega_{q_0}\right|}{2 B_{q_0}(N+1)}$. It shows a good coincidence with the numerical results (Fig.\ref{fig6}). Note, that increasing the chain length decreases the instability threshold and increases its variation. Thus, in sufficiently large arrays the solution will loose stability at very small nonlinearities with the probability, almost equal to that of $\Delta V_{q_0,q_0}$ being negative, which is $0.5$ in our case.

If the instability depends that sensitively on the disorder realization, there must be certain classes of inhomogeneities that augment it or suppress. Identifying them offers the possibility of controlling the energy flow in the mode space by designing specific impurities and, further, changing them in time. The disorder induced correction in (\ref{eq11}) reduces to 
%\begin{equation}\label{eq11a}
%\begin{aligned}
%&
$\Delta V_{q_0,q_0} =\frac{8}{\sqrt{N+1}}\sum\limits_{n=1}^{N-1}\kappa_n\left(\cos{\frac{\pi q_0(2n+1)}{N+1}}\sin^2{\frac{\pi(2n+1)}{2(N+1)}}+O(N^{-2})\right)$.
%\end{aligned}
%\end{equation}
Note, that it is linear in $\kappa_n$, and, therefore, one can represent $\kappa_n$ as a sum of spatial Fourier components, their contributions being additive to $\Delta V_{q_0,q_0}$. 

Let us consider a harmonic inhomogeneity $\kappa_n=\frac{1}{2}\cos{(\frac{\pi p (n+1/2)}{N+1}+\varphi)}$, where $\varphi$ is the phase shift. It is natural to expect the absolute extrema of $\Delta V_{q_0,q_0}(p,\varphi)$ (and the maximal gain or loss in stability) to be reached for $p=2q_0$. This case yields $\Delta V_{q_0,q_0}\approx\sqrt{N+1}\cos{\varphi}$. Thus, the bifurcation point reaches its maximum and minimum for $\varphi=\pi$ and $\varphi=0$ respectively, giving $\mu^*\approx\mu_0^*\left(1\pm\frac{D(N+1)^2}{\pi^2\left|\Omega_{q_0}\right|}\right)$. At the same time one gets a zero shift for $\varphi=\pm\frac{\pi}{2}$. Analogously, for $p=q_0\pm1$ one gets $\mu^*\approx\mu_0^*\left(1\pm\frac{8 D(N+1)^2}{3\pi^3\left|\Omega_{q_0}\right|}\sin{\varphi}\right)$, which is $-\frac{\pi}{2}$ shifted in $\varphi$ and has a bit smaller amplitude. For $p=q_0\pm2$ it reads $\mu^*\approx\mu_0^*\left(1+\frac{D(N+1)^2}{2\pi^2\left|\Omega_{q_0}\right|}\cos{\varphi}\right)$, which is $\pi$ shifted in $\varphi$ and has twice a smaller amplitude. Larger deviations from $2q_0$ lead to progressively decreasing shifts.

\begin{figure}[t]
{\centering
\resizebox*{0.95\columnwidth}{!}{\includegraphics{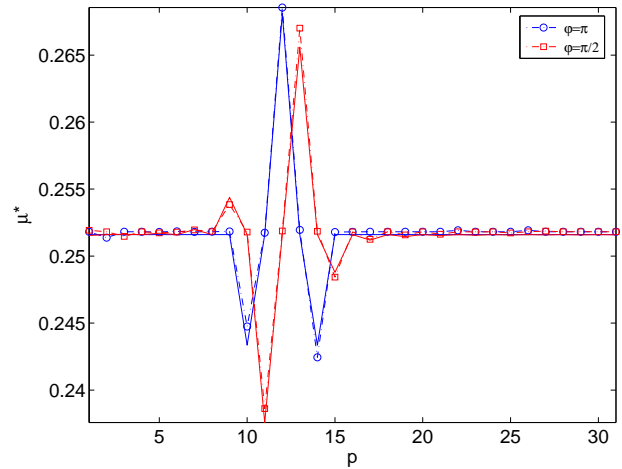}}}
{\caption{The instability threshold for QBs with $q_0=10, N=32$ and $\kappa_n=\frac{1}{2}\cos{\left(\frac{\pi p (n+1/2)}{N+1}+\varphi\right)}$. Dash-dotted and marked lines are numerical results for (\ref{eq6a}), solid lines are analytical estimates}\label{fig7}}
\end{figure}

These results are illustrated in Fig.\ref{fig7}, and show a good correspondence to the numerically determined QB stability. Summing up, the spatial Foirier components $p\in[2q_0-2,2q_0+2]$ of $\{\kappa_n\}$ are decisive for the $q_0$-QB stability. The dependence is notably different and much more complicated than the possible "na\"{i}ve" expectation that harmonic inhomogeneities with $p=q_0$ will most effectively stabilize or destabilize $q_0$-QBs. A remarkable fact is the sensitive dependence on the phase of the impurity harmonics: even for a fixed $p$ opposite shifts in the threshold occur. Presumably, this is the consequence of the deformation of the linear spectrum due to inhomogeneities, as $\Delta V_{q_0,q_0}$ is, actually, the difference in the frequency shifts of linear modes (\ref{eq10c}). In its turn, this is determined by the boundary conditions, which also affect the nonlinearity induced interaction. It clearly highlights one of the future directions of study.

These findings suggest a possibility of controlling the energy flow between modes. Indeed, by imposing a proper spatially periodic modulation of the linear coupling one can destabilize certain QB exictations and (i) speed up equipartition or (ii) stabilize others, where the energy will be radiated to and trapped. New QBs may also be destabilized to arrange a further energy flow. Experimentally, in miniature mechanical lattices inhomogeneities can be created, for example, by laser heating, either as harmonic or spot impurities, like it was designed to control discrete breathers relocation in cantilever arrays \cite{Sievers}. In optical lattices one can implement the same technique that has been recently used for generating disordered potentials in studies of the Anderson localization of light \cite{segev} and matter (BEC) \cite{lahini} waves. 

In summary, we have generalized the concept of QBs to the case of non-zero disorder and analyzed these nonlinear vibrational modes in weakly disordered DNLS arrays. We demonstrated, that QBs remain exponentially localized in the mode space and stable, if the disorder is sufficiently small. Their stability depends sensitively on a particular realization of disorder, and may be enhanced or undermined. The prevailing contribution to the stability is made by the spatial harmonics of disorder which wave numbers are close to twice of that of the QB seed mode. Thus, inhomogeneities design appears to be a promising technique of controlling the energy flow between nonlinear modes. We expect these ideas and methods to apply to a variety of nonlinear weakly disordered lattices -- and have already demonstrated it for the FPU chain (to be reported elsewhere) -- to name the disorder coming from other sources (masses, nonlinearities), higher dimensions, and quantum arrays. We believe that the results on the nonlinear modes sustainability, stability, and controlling will be widely adopted in experiments and applications.

We thank S. Flach for stimulating discussions.% and O. Kanakov for useful comments.

\end{document}